\documentclass[pre,twocolumn,aps,superscriptaddress,showpacs,floatfix]{revtex4}
\usepackage{graphicx}
\usepackage{dcolumn}
\usepackage{bm}
\usepackage{amssymb,amsmath}
\usepackage[usenames,dvipsnames]{color}

\begin{document}
\title{Coarsening modes of clusters of aggregating particles}
\author{Andrey Pototsky}
\affiliation{Mathematics Discipline, Faculty of Engineering and Industrial sciences, Swinburne University of Technology, Hawthorn, Victoria, 3122, Australia}
\author{Uwe Thiele}
\affiliation{Department of Mathematical Science, Loughborough University, Loughborough LE11 3TU, United Kingdom}
\affiliation{Institut f\"ur Theoretische Physik, Universit\"at M\"unster, Germany}
\author{Andrew J. Archer}
\affiliation{Department of Mathematical Science, Loughborough University, Loughborough LE11 3TU, United Kingdom}

\begin{abstract}
There are two modes by which clusters of aggregating particles can coalesce: The clusters can merge either (i) by the Ostwald ripening process in which particles diffuse from one cluster to the other whilst the cluster centres remain stationary, or (ii) by means of a cluster translation mode, in which the clusters move towards each other and join. To understand in detail the interplay between these different modes, we study a model system of hard particles with an additional attraction between them. The particles diffuse along narrow channels with smooth or periodically corrugated walls, so that the system may be treated as one-dimensional. When the attraction between the particles is strong enough, they aggregate to form clusters. The channel potential influences whether clusters can move easily or not through the system and can prevent cluster motion. We use Dynamical Density Functional theory to study the dynamics of the aggregation process,  focusing in particular on the coalescence of two equal size clusters. As long as the particle hard-core diameter is non-zero, we find that the coalescence process can be halted by a sufficiently strong corrugation potential. The period of the potential determines the size of the final stable clusters. For the case of smooth channel walls, we demonstrate that there is a cross-over in the dominance of the two different coarsening modes, that depends on the strength of the attraction between particles, the cluster sizes and the separation distance between clusters.
\end{abstract}
%
%
\maketitle
\section{Introduction}
The late-stage dynamics of phase separation is often characterised by
the presence of droplets, islands or clusters of one phase dispersed in a
surrounding medium of another phase. This may for example be: droplets
of a liquid condensing from a supersaturated vapour, phase separation
in a binary mixture or alloy, or the condensation to form islands of a
liquid or solid phase as particles diffuse over a surface. Over time
the average size of the clusters or droplets grows and the total number
decreases. There are two mechanisms by which this can occur. {The first
is often referred to as Ostwald ripening, in which particles diffuse
from the smaller clusters to the larger clusters through the
surrounding phase while the centres of the clusters remain fixed in
space. Lifshitz and Slyozov \cite{LiSl61} and Wagner \cite{Wagner61}
were the first to develop a theory for this process. For a more recent
discussion see for example Refs.~\cite{Onuki, DeKa09, MaRo84,
  Voorhees85, YEGG93, bray94}.  The second mechanism by which the
clusters may coarsen is via the motion of entire clusters towards
other clusters and subsequently joining. This motion may simply be via
droplet diffusion \cite{Onuki}. Alternatively, it can be due to an
imposed external driving, e.g.~by gravity in sedimentation, creaming and the formation of rain drops.
Quite often cluster diffusion may be neglected compared to the diffusion
of individual particles since the diffusion coefficient $D$ through the
surrounding material decreases with the inverse of the radius $R$ of the
diffusing object, e.g., in three dimensions $D\sim 1/R$.}
Since the radius of a cluster is much larger than an individual
particle, the cluster diffusion coefficient is much smaller. However,
this is not always the case, particularly when two drops come close
enough to each other that they begin to interact strongly. Then
capillary and other such forces can lead to their motion and may
result in cluster-cluster coalescence.

There are other aspects and theories of aggregation kinetics that are worth mentioning: In earlier models the aggregation process was regarded as irreversible, whereby the thermally activated escape of single particles from clusters was neglected. In this regime the growth occurs solely due to collisions between smaller clusters that encounter each other traveling by diffusion. This aggregation scenario is referred to as the Diffusion-Limited or Reaction-Limited aggregation \cite{meakin83,kolb83,weitz84,lin89}. Later models assume an extended aggregation scenario by taking into account the reversible reorganisation processes, associated with the breaking away and diffusing of particles from one cluster to another \cite{botet85,filoche00,grosskinsky02}.

In reality, all of the above mechanisms play a part in the coarsening
dynamics. {However, in different regimes different mechanisms
  dominate.} When the distance between clusters is large, then the
Ostwald ripening mechanism dominates, but the escape of particles from
clusters can be an extremely slow process. {However, when the
  density of clusters is sufficiently high, they interact and cluster
  motion can dominate. Note too that the Ostwald ripening mechanism
  depends on the solubility of the clustering particles in the
  background (dispersing) phase, implying that for strongly phase
  separating systems, the Ostwald coarsening can be very slow.} The
cluster motion process also depends strongly on the properties of the
dispersing phase and also in the case of systems of particles moving
over a surface, on the nature of the interactions with the surface and the degree of surface roughness, since clusters may become pinned on a rough surface. In this paper we develop a simple model to study the interplay of the different coarsening processes and the influence of the background substrate.

The model we consider here consists of system of one dimensional (1D)
Brownian particles (i.e.\ rods) confined in a periodic external
potential $U(x)$. Such a situation occurs when particles are confined
within a {narrow pore, groove or channel.} The amplitude of the
modulations in $U(x)$ has a significant influence on the rate at which
individual particles and also clusters can move through the
system. The particles are assumed to have a hard core of diameter $\sigma$
and also an attractive interaction. This system also serves as a
minimal model of molecular transport through micro- and nano-pores
featuring the so-called single file diffusion \cite{lutz04}. We focus,
in particular, on the case of two clusters in the system and study in
detail the dynamics of how these clusters join together. We also focus
on small clusters, where it is easier to understand the processes that occur during coalescence. However, we relate our results to larger clusters and we believe our results are quite general.

We find that the Ostwald and translation aggregation mechanisms
discussed above are both exhibited by our model system, depending on
the values of the system parameters and we also find that as these are
varied, there can be a crossover from one mode dominating to the
other. The interplay of these two modes was studied previously for
{shallow liquid droplets on homogeneous and heterogeneous
  surfaces \cite{TBBB03,GlWi2003pre,PiPo2004pf,GlWi2005pd,Thie2007}.  There, the
  analogue of the Ostwald process is referred to as the mass transfer
  mode of the coarsening dynamics of dewetting liquid films
  \cite{TBBB03} or the mass exchange and collapse mode
  \cite{GlWi2005pd}. This refers to when one of two neighbouring liquid
  droplets shrinks and the other one grows, leading eventually to a
  single large droplet. Similarly, there is a translation mode
  \cite{TBBB03} called the spatial motion and collisions mode in
  \cite{GlWi2005pd}, whereby neighbouring droplets move together and
  join. We see here that there are strong parallels between the
  aggregation dynamics of colloidal (Brownian) systems and the
  coarsening dynamics in binary mixtures or in dewetting liquid films
  \cite{bray94,TBBB03,GlWi2005pd}. The qualitative behaviour of the
  present model is very similar to that observed in these systems and
  so we believe that the overall qualitative conclusions we draw in
  this paper are rather more general than one might expect, given the
  simple 1D model that we study. This is due to the fact that the
  coarsening modes are closely related to the symmetry modes of the
  systems \cite{Thie2007}.}

A particular application of our results is the aggregation in colloidal
systems, since we assume the particles move by Brownian dynamics. The
rise of nanotechnology has seen a surge of interest in aggregation and cluster formation of colloidal particles with nano-clusters hailed as prototypes of novel functionalized materials \cite{manoharan03, stradner04}. The formation of clusters of attracting colloids with their subsequent aggregation into a new solid-like phase is closely linked to the phenomena of glass transitions and gelation \cite{segre01, campbell05}. At a given volume fraction, the uniform density state, which corresponds to a homogeneous suspension of colloids, can become unstable if a certain critical attraction strength is reached. As the system evolves towards a global minimum of the free energy, the initially formed clusters merge to form bigger aggregates. The final state, which is found in the long time limit, crucially depends on the nature of the interaction potential between the particles. Typically, if the interaction is purely attractive up to the hard-core cut-off, the global minimum of the free energy corresponds to a single infinitely large aggregate. The process when large colloidal clusters collapse to form a disordered solid is known as gelation \cite{dawson02, weitz84, campbell05}. 

To describe the dynamics of cluster aggregation we take a statistical
mechanical point of view and study the kinetics of the aggregation
process using dynamical density functional theory (DDFT)
\cite{marini99, marini00, ArEv04, ArRa04}, which is a theory for the
time evolution of the one body density distribution $\rho(x,t)$ of the
particles in the system. DDFT builds on equilibrium density functional
theory \cite{evans, evans1992fif, HM} and therefore in principle
guarantees giving the correct equilibrium fluid density profile, even
though its description of the dynamics involves approximation
\cite{marini99, marini00, ArEv04, ArRa04}. An advantage of DDFT over
some other approaches, such as the previously used discrete models
\cite{meakin83, kolb83, weitz84, lin89, botet85, filoche00,
  grosskinsky02} is that DDFT takes as input the microscopic
properties of the particle interactions and particle dynamics. The
approximation for the equilibrium Helmholtz free energy functional
that we use is based on the exact free energy functional for a 1D
fluid of hard rods \cite{percus76}. The contribution to the free
energy due to the attraction between the particles is then treated
using a simple mean field approximation \cite{evans1992fif}. This is
based on the DDFT for aggregating hard rods transported along narrow
channels developed in Refs.~\cite{PA10, PA11}, where the model is used to study the directed transport of particles through channels with periodically corrugated walls.

The paper is structured as follows: In Section~II we give a brief
overview of the DDFT for attractive hard rods and discuss the bulk
system phase behaviour, in particular recalling {the aspects of the
linear stability analysis of the homogeneous phase of the model that
are relevant to the present study.} In Section III, we discuss how to
obtain the eigenfunctions and eigenvalues corresponding to the two
different coarsening modes discussed above and illustrate this with
some typical results. We show that cluster coarsening can be halted by
applying a sufficiently strong pinning potential $U(x)$. The critical
strength of the potential, which is necessary to stop clustering,
greatly depends on the diameter of the particle $\sigma$ and in fact
diverges in the limit $\sigma\to0$. We show that for small and intermediate
amplitudes of the external potential the aggregation process is
dominated by the Ostwald mode if the attraction between particles is
rather weak. However, as the attraction strength is increased, the
translation mode becomes dominant, whereas the Ostwald mode becomes
inactive. In Section IV we discuss further {how  a cross-over
  from one coarsening mode to the other can occur}. Finally, in Section V we make a few concluding remarks.

\section{Model system and DDFT}
Following \cite{PA10,PA11}, we consider a system of $N$ hard particles of diameter $\sigma$, moving in a channel of length $S$. We impose periodic boundary conditions -- i.e.\ the channel may be considered to be circular. The channel is assumed to be sufficiently narrow that the particles are confined to a line and so may be treated as a 1D system of hard rods. The particles interact with the corrugated channel walls through a periodic potential $U(x)$, which is of period $L=S/n$, where $n$ is an integer. In addition to the hard core repulsion between the particles, there is an attractive interaction between the particles that is described by the potential $w(x) = -\alpha\exp{(-\lambda x)}$, with the typical interaction range $1/\lambda$ and strength $\alpha>0$. We assume that the dynamics of the particles are governed by over damped stochastic equation of motion and so we may use DDFT \cite{marini99, marini00, ArEv04, ArRa04} to approximate the Fokker-Planck equation for the time evolution of the  one-body density distribution $\rho(x,t)$ as follows
\begin{eqnarray}
\label{FP}
\frac{1}{\Gamma}\frac{\partial \rho(x,t)}{\partial t} = \frac{\partial}{\partial x}\left[ \rho(x,t)\frac{\partial }{\partial x}\frac{\delta F[\rho(x,t)]}{\delta \rho(x,t)}\right],
\end{eqnarray}
where $\Gamma$ is the mobility of a single particle and the Helmholtz free energy functional $F[\rho]$ is given by
\begin{eqnarray}
\label{dft_energy}
F[\rho(x,t)] &=& k_B T \int_{-\frac{S}{2}}^{\frac{S}{2}} dx\,\rho(x,t)[\ln{\Lambda\rho(x,t)}-1] \nonumber \\
&+& \int_{-\frac{S}{2}}^{\frac{S}{2}} dx\,U(x,t)\rho(x,t) \nonumber \\
&+&  F_{\rm hc}[\rho] + F_{\rm at}[\rho].
\end{eqnarray}
In Eq.\,(\ref{dft_energy}) the first term is the ideal-gas contribution to the free energy, where $\Lambda$ is the thermal de Broglie wavelength, $k_B$ is Boltzmann's constant and $T$ is the temperature. The term $F_{\rm at}$ is the excess free energy contribution due to the attraction between particles, for which we use the following mean-field approximation \cite{evans1992fif}:
\begin{equation}
\label{eq:F_at}
F_{\rm at}[\rho] = \frac{1}{2}\int_{-\frac{S}{2}}^{\frac{S}{2}} dx \int_{x-\frac{S}{2}}^{x+\frac{S}{2}} dx'\,w(\mid x-x' \mid)\rho(x)\rho(x').
\end{equation}
The term $F_{\rm hc}[\rho]$ in Eq.\,(\ref{dft_energy}) is the free energy contribution from the hard-core interaction between the particles. We use the Percus exact functional \cite{percus76} for 1D hard rods of length $\sigma$:
\begin{eqnarray}
\label{hr_energy}
F_{\rm hc}[\rho] = \frac{1}{2}\int_{-\frac{S}{2}}^{\frac{S}{2}} dx\, \phi[\rho(x)] \left\{\rho\left(x+\frac{\sigma}{2}\right)+\rho\left(x-\frac{\sigma}{2}\right)\right\},
\end{eqnarray}
where $\phi[\rho(x,t)] = -k_B T\ln{[1-\eta(x,t)]}$ and $\eta(x,t) = \int_{x-\sigma/2}^{x+\sigma/2}dx'\,\rho(x',t)$. 

Any solution of Eq.\,(\ref{FP}) is subject to periodic boundary conditions in the interval $[-S/2,S/2]$ and the normalization condition $\int_{-S/2}^{S/2}\rho(x)\,dx=N$. 

To non-dimensionalize Eq.\,(\ref{FP}) we rescale $x$ with the characteristic length of the attraction potential $(\lambda)^{-1}$, time $t$ with $(\lambda)^{-2}/\Gamma k_B T$, which represents the characteristic time for a particle to diffuse a distance $(\lambda)^{-1}$, and all energy units are scaled with $k_B T$.  For simplicity we use the same notations for the rescaled units, implying that non-dimensional Eqs.\,(\ref{FP}) -- (\ref{hr_energy}) are obtained by setting $\Gamma=1$, $\lambda=1$ and $k_B T=1$.

We assume for simplicity that the periodic external potential has the following form
\begin{eqnarray}
\label{potential}
U(x)=-\frac{\chi}{2\pi}\cos{\left(\frac{2\pi x}{L}\right)}.
\end{eqnarray}
The amplitude parameter $\chi$ governs the height of the potential
barrier between neighbouring minima in the potential and $L$ is the
period of the corrugation potential exerted onto the particles by the
channel walls. Our simple 1D model may also be used to understand the
collective dynamics or particles diffusing over a rough surface. In
this context, $\chi$ may be thought of as a parameter that
characterises the roughness of the surface. We also study smooth
channels, which correspond to the limit $\chi\to0$. In this case, when
$U(x)=0$, if the particles condense to form a cluster or several
clusters and so are non-uniformly distributed with density profile
$\rho(x)$, then in this situation the free energy of the system does
not change if there is an arbitrary translation in the system -- i.e.\
when $\rho(x) \to \rho(x+l)$, where $l$ is the arbitrary shift
distance. {In the numerical treatment of this case some care has to be
taken because of the translation symmetry (see next section).}
\section{Phase behaviour and stability of the uniform density state}
In Ref.~\cite{PA11} the equilibrium phase behaviour of the model
system defined in Eqs.\,(\ref{FP}) -- (\ref{hr_energy}) is studied. As
the attraction strength $\alpha$ is increased (or, equivalently, as
the temperature $T$ is decreased) {the uniform density state
$\rho(x)=\rho_0$ that exists for $\chi=0$ becomes unstable for some
range of densities $\rho_0$, and phase separation occurs.} Whilst
there is no true phase transition in the system as predicted by the
present mean-field model because the system is one dimensional (i.e.\
fluctuations round the predicted transition), nonetheless, a
comparison with Brownian dynamics computer simulations reveals that
the DDFT (\ref{FP}) -- (\ref{hr_energy}) does describe well the
aggregation of particles into clusters \cite{PA11}. A cluster state
may form spontaneously, if the uniform state is linearly unstable, or
there may be a free energy barrier that must be overcome to form
clusters -- i.e.\ clusters must be nucleated \cite{Onuki, PA11}. A
linear stability analysis \cite{PA11, ArEv04} shows that the trivial
steady state solution $\rho(x)=\rho_0$ of Eq.\,(\ref{FP}), can become linearly unstable with respect to two distinct instability modes: the spinodal mode and the freezing mode, depending on the value of $\rho_0$. Expanding on the right hand side of Eq.~\eqref{FP} in powers of the density, one obtains an equation of the form
\begin{eqnarray}
\label{expand_FP}
\partial_t\rho=\hat{L}\rho+{\cal O}(\rho^2),
\end{eqnarray}
where $\hat{L}(x)$ is a linear operator. In the linear stability
analysis the density is assumed to be of the form
$\rho(x,t)=\rho_0+\epsilon \exp(\beta(k)t+ikx)$, where the amplitude
$\epsilon$ is assumed to be small. Substituting this into Eq.\
\eqref{expand_FP} and then linearising in $\epsilon$, one obtains
 {the dimensionless dispersion relation \cite{PA11}
\begin{eqnarray}
\label{disp_rel}
\beta(k)=&-&k^2-\frac{2\rho_0k\sin(k\sigma)}{1-\rho_0\sigma}\nonumber\\
&-&\frac{4\rho_0^2\sin^2(k\sigma/2)}{(1-\rho_0\sigma)^2}+\frac{2\alpha\rho_0k^2}{k^2+1}.
\end{eqnarray}
}
In the case of an infinitely extended system $(S\rightarrow \infty)$,
the spinodal instability sets in at zero wave number $k=0$.  {In contrast,
the freezing instability sets in with a finite critical wave number
$k=k_c\neq 0$.} The linear stability analysis allows for a
determination of the regions of the phase diagram where phase
separation occurs and identifies the length scales of the structures
that are formed in the initial stages after a uniform state is
quenched into a linearly unstable parameter region \cite{PA11,
  ArEv04}. However, here our focus is {not on the linear
  short-time dynamics, but on the long-time dynamics,} when clusters
are already present, whether formed by nucleation or spontaneously
after a quench. These clusters coarsen over time, via one or both of
the two mechanisms described in the introduction: either the Ostwald
ripening mechanism, or via the motion and coalescence of clusters. 

To calculate solution branches of steady cluster density profiles and
their bifurcations numerically, we employ the method presented in
Ref.~\cite{PA11}. Namely, we rewrite our governing
integro-differential equations as a finite dimensional dynamical
system, i.e.~as a finite set of ordinary differential equations in
time. To do so, Eq.\,(\ref{FP}) is Fourier-transformed truncating at a
certain number of modes $M_f$, resulting for our gradient dynamics
Eq.\,(\ref{FP}) in $M_f$ real equations for the Fourier coefficients
that form the dynamical system. Throughout the paper we set
$M_f=400$. Its solution and bifurcation behaviour is systematically
analysed combining fast Fourier transforms with pseudo-arclength path
continuation \cite{DoKK1991ijbc,Kuzn10,Dijk14} as bundled in the
package auto07p \cite{AUTO}.

The continuation may either be started at a steady state obtained via
a direct time simulation or from analytically known small-amplitude
harmonic solutions that well approximate non-uniform states close to
the primary bifurcations of the uniform state \cite{PA11}.  However,
to numerically obtain $S$-periodic steady state solutions of
Eq.\,(\ref{FP}) in a smooth channel, i.e.\ the limit $\chi\to0$, one
has to overcome a problem which is related to the translational
invariance of solutions in this case.  In practice, we use the
homotopy method to obtain a weakly modulated steady state with a
period equal to the system size, i.e., we set $L=S$. We start the
continuation with a constant density $\rho(x)=\rho_0$, which is a
steady state solution when $\alpha=\sigma=0$ (so that $F_{\rm
  at}=F_{\rm hc}=0$) and $U(x)=0$.  We then introduce an additional
homotopy parameter, which multiplies the last three terms in
Eq.\,(\ref{dft_energy}). In the first continuation run, the homotopy
parameter is increased from zero to one keeping $\chi$, $\alpha$ and
$\sigma$ at generic values in the linearly stable parameter
region. This results in a weakly modulated steady state. In the second
run, this steady state is continued as $\alpha$ is increased beyond
the point where condensation sets in and large amplitude cluster
solutions emerge.  In the next run, this large amplitude solution is
continued when the heterogeneity amplitude $\chi$ is decreased until
it becomes vanishingly small (typically $\chi=10^{-5}$ well
approximates a smooth channel). Thus, the potential $U(x)$ of nearly
vanishing amplitude serves in this situation merely as a numerical
trick to break the degeneracy related to translational invariance and
to fix the steady state solutions in space.

\section{Coarsening of clusters}

In this section we focus on the dynamics of clusters that have been
created either by means of nucleation from a metastable uniform state,
or by the linear instability (spinodal or freezing). In particular, we
consider the coalescence of two clusters into one bigger aggregate. Note, that clusters can also be created in the absence of the attraction between the particles, by the potential $U(x)$. In this case the particles form clusters due to accumulation around the local minima of $U(x)$.

\subsection{Symmetry and coarsening modes}
Any steady state solution of Eq.\,(\ref{FP}) in a smooth channel,
i.e.\ with $U(x)=0$, has two types of symmetry: (i) a translation
symmetry and (ii) a volume change symmetry. The existence of these two
symmetries for individual clusters gives rise to the two
  different coarsening mechanisms discussed above if they are combined
  for several clusters. This is sketched for droplets on a solid
  substrate in Ref.~\cite{Thie2007}.

For the translation symmetry one sees that if $\rho(x)$ is a
single-cluster steady state solution of Eq.\,(\ref{FP}), then
$\rho(x+\xi)$ with an arbitrary shift $\xi$ is also a solution. For
small $\xi$ one can write $\rho(x+\xi)\approx \rho(x)+\xi \partial_x
\rho(x)$, which implies that $h(x)=\partial_x \rho(x)$ is the
eigenfunction of the DDFT operator $\hat{L}(x)$ linearized around the
steady state $\rho(x)$, with zero eigenvalue $\gamma=0$,
i.e. $\hat{L}(x) \partial_x \rho(x) =0$. In what follows we refer to
$h(x)=\partial_x \rho(x)$ as the translation symmetry mode
eigenfunction, which is sometimes called Goldstone mode of this
continuous symmetry.  Combining translation modes with opposite signs
for two adjacent droplets gives the translation mode of coarsening.

The volume change symmetry is associated with the normalization
(particle conservation) condition of the solution of
Eq.\,(\ref{FP}). Given a density profile $\rho(x)$, which is a single-cluster steady
state solution of Eq.\,(\ref{FP}), with
\begin{equation}
\label{norm}
\frac{1}{S}\int_{-S/2}^{S/2}\rho(x)\,dx=\rho_0,
\end{equation}
we now consider the density profile $\tilde{\rho}(x)=\rho(x)+\delta
\rho(x)$, with
$(1/S)\int_{-S/2}^{S/2}\tilde{\rho}(x)\,dx=\tilde{\rho}_0=\rho_0 +
\delta \rho_0$ that is also a steady state solution of Eq.\,(\ref{FP})
for some small $\delta \rho_0$. Since $\tilde{\rho}(x)=\rho(x)+\delta \rho(x)$, we deduce that $h(x)=\delta \rho(x)$ {is another eigenfunction of the operator $\hat{L}(x)$ with zero eigenvalue $\gamma=0$}, i.e. $\hat{L}(x) \delta \rho(x) =0$. Consequently, the volume symmetry mode $\delta \rho(x)$, corresponds to the derivative with respect to the density $\rho_0$, i.e. $\delta \rho(x) = \delta \rho_0 \partial \rho(x) / \partial \rho_0$, where we define
\begin{equation}\label{rho_deriv}
\frac{\partial\rho(x)}{\partial\rho_0}\equiv\lim_{\delta\rho_0\to0}\frac{\rho(x;\rho_0+\delta\rho_0)-\rho(x;\rho_0)}{\delta\rho_0}.
\end{equation}
Combining volume modes with opposite signs for two adjacent
  droplets gives the mass transfer mode of coarsening which we refer
  to as the Ostwald ripening mode. Note that one may alternatively
calculate the Ostwald mode eigenfunction by considering the change
$\delta\rho(x)$ from an infinitesimally small change in the value of
the chemical potential, $\mu_0\to\mu_0+\delta\mu_0$.

Thus, any steady state solution $\rho(x)$ has the two neutrally
  stable symmetry modes. If these respective modes are combined for
  pairs or groups of clusters one obtains the various coarsening mode eigenfunctions
that for homogeneous systems normally have small positive eigenvalues
resulting from the overlap of the tails of the symmetry modes of the
individual clusters.
If one now breaks the translation symmetry by considering a
heterogeneous system with $U(x)\neq0$, the stability and
characteristics of translation and Ostwald coarsening modes closely
related to the two introduced eigenfunctions still define the dynamics
of coarsening.

In order to illustrate this discussion, one may examine the simple
situation where an ideal gas of non-interacting particles (i.e.\ $\sigma=0$
and $\alpha=0$) is confined within a parabolic potential $U(x)=ax^2$,
where $a$ determines the strength of the confinement. This is arguably
the simplest example of a cluster, where the particles remain in the
cluster due to the confinement of the external potential. The
equilibrium density profile is obtained from minimising the the grand
potential \cite{evans1992fif}
\begin{equation}
\Omega[\rho]=F[\rho]-\mu\int_{-S/2}^{S/2}\rho(x)dx,
\end{equation}
where the Helmholtz free energy $F$ results from
Eq.~\eqref{dft_energy} and $\mu$ is the chemical potential, which is
the Lagrange multiplier enforcing the constraint \eqref{norm}. When
$S$ is large, we obtain the density profile
$\rho(x)=\rho_0\exp(-ax^2)$, where $\rho_0=\Lambda^{-1}\exp(\mu)$.
Thus, the translation mode eigenfunction is
$h(x)=\partial_x\rho(x)=-2xa\rho_0\exp(-ax^2)$. In order to determine
the volume mode eigenfunction we note that
$\rho(x;\rho_0+\delta\rho_0)=(\rho_0+\delta\rho_0)\exp(-ax^2)$, and so
using Eq.\,\eqref{rho_deriv} we find that the volume mode
eigenfunction is $h(x)=\delta \rho(x) / \delta
\rho_0=\exp(-ax^2)$. Note that the translation and volume mode
eigenfunctions are orthogonal. We now consider two such confining
potentials generating two clusters a distance $l$ apart. In this
situation $U(x)=a(x+l/2)^2$, for $x<0$ and $U(x)=a(x-l/2)^2$, for
$x>0$. The resulting density profile exhibits two `clusters' centered
at $\pm l$ and the density profile
\begin{eqnarray}\label{rho_andy}
\rho(x)= 
\begin{cases}
\rho_0\exp(-a(x+l/2)^2) \qquad\mbox{for}\quad x<0\\
\rho_0\exp(-a(x-l/2)^2)\qquad\mbox{for}\quad x>0,
\end{cases}
\end{eqnarray}
Suppose now that there are slightly more
particles ($\delta \rho_0$) in the left
hand cluster than in the right hand cluster. In this situation, the
Ostwald mode eigenfunction is:
\begin{eqnarray}\label{Ostwald}
h(x)\approx 
\begin{cases}
\exp(-a(x+l/2)^2) \qquad\mbox{for}\quad x<0\\
-\exp(-a(x-l/2)^2)\qquad\mbox{for}\quad x>0,
\end{cases}
\end{eqnarray}
This eigenfunction indicates that there will be a diffusion of
particles from one side to the other. One must calculate the
eigenvalue to determine the direction of this flux, although in
  this simple situation one knows already that the symmetric state is
  stable, i.e., the eigenvalue must be negative and the flux is from
  left to right.  One may also examine the translation mode
eigenfunction, although in this simple case this mode is not
relevant. However, when we introduce the attraction between the
particles, this mode can become relevant and the two clusters can seek
to move toward one another.

\begin{figure}[t]
\centering
\includegraphics[width=0.99\columnwidth]{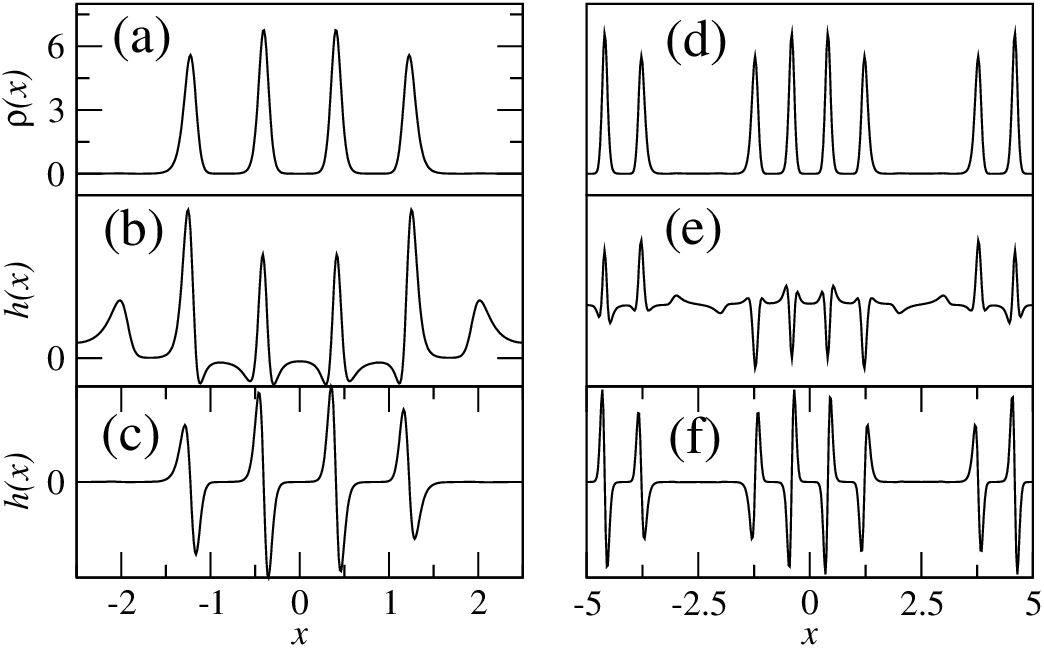}
\caption{
  \label{F4a} (a) {Density profile for a single cluster
    which is a period-$L$ steady state solution in the smooth channel
    ($U(x)=0$) for $\rho_0=0.8$, $\alpha=15$, $\sigma=0.75$ and domain size
    $S=L=5$. (b) The volume symmetry mode and (c) translation symmetry mode eigenfunctions
    $h(x)$ for a single cluster with eigenvalues $\gamma=0$. (d)
    Density profile for a two cluster state, i.e., Period-$L$ steady
    state solution in domain $S=2L=10$ for parameters as in (a). (e)
    Ostwald ripening coarsening mode with $\gamma=-0.8$ (stable). (f)
    Translation coarsening mode with $\gamma=3.5$ (unstable). }
}
\end{figure}

We now consider a small system of strongly interacting particles. We
display in Fig.~\ref{F4a}(a) {a single four-particle cluster,
  i.e.\ a period-$L$ steady state density profile obtained for
  $\sigma=0.75$, $\alpha=15$, $\rho_0=0.8$ and $S=L=5$. Due to the very
  strong attraction, the four particles are squeezed into the compact
  cluster with four distinct density peaks. Figs.~\ref{F4a}(b) and (c)
  show the volume symmetry mode and the translation symmetry mode
  eigenfunctions of the steady state, respectively.}

{If $L$ is the period of a (neutrally) stable state $\rho(x)$,
  then $\rho(x)$ is also a steady state periodic solution in a system
  with the larger size $S=nL$, with an arbitrary integer $n$.  In such
  an extended system, the solution $\rho(x)$ is still (neutrally)
  stable with respect to perturbations with period $L$, but may be
  unstable with respect to perturbations with period $nL/m$,
  $(m=1,\dots,n)$ that break the internal discrete symmetry of
  $\rho(x)$ under translations by $L$, i.e., to coarsening modes.}

Here we are particularly interested in the case of $n=2$, i.e.\ when
$S=2L$. We refer to the period-$L$ steady state $\rho(x)$ as a cluster
of the length $L$. Then, the case $n=2$ corresponds to two identical
clusters of period-$L$ in a system of total size
$S=2L$. Fig.\,\ref{F4a}(d) shows the density profile for such a two cluster state, i.e., 
two period-$L$ ($L=5$) clusters of four particles each in a system of
size $S=2L=10$. In the following sections we describe the coalescence
dynamics of the two period-$L$ clusters that are stable with respect
to perturbations with wavelength $L$, but may be unstable with respect
to perturbations with wavelength $2L$. These (stable or unstable)
period-$2L$ perturbations are the two coarsening modes discussed
above. The mode associated with the Ostwald ripening coarsening process can be approximated by combining the volume modes of two individual clusters taken with opposite signs, as shown in Fig.\,\ref{F4a}(e). See also Eq.~\eqref{Ostwald}. The second mode is the translation coarsening mode, which is obtained by combining the translation modes of the individual clusters taken with opposite signs. This is shown in Fig.\,\ref{F4a}(f). For the case in Fig.\,\ref{F4a} the eigenvalue for the Ostwald coarsening mode $\gamma=-0.8$, is negative and so the system is stable against this mode. However, for the translation mode $\gamma=3.5$ and so the system is unstable against this mode. Because this mode is unstable, the two clusters move towards each other, preserving their shape until they collide to form a bigger cluster.

\subsection{Purely attracting particles}
\label{secIIIB}

\begin{figure}[t]
\centering
\includegraphics[width=0.99\columnwidth]{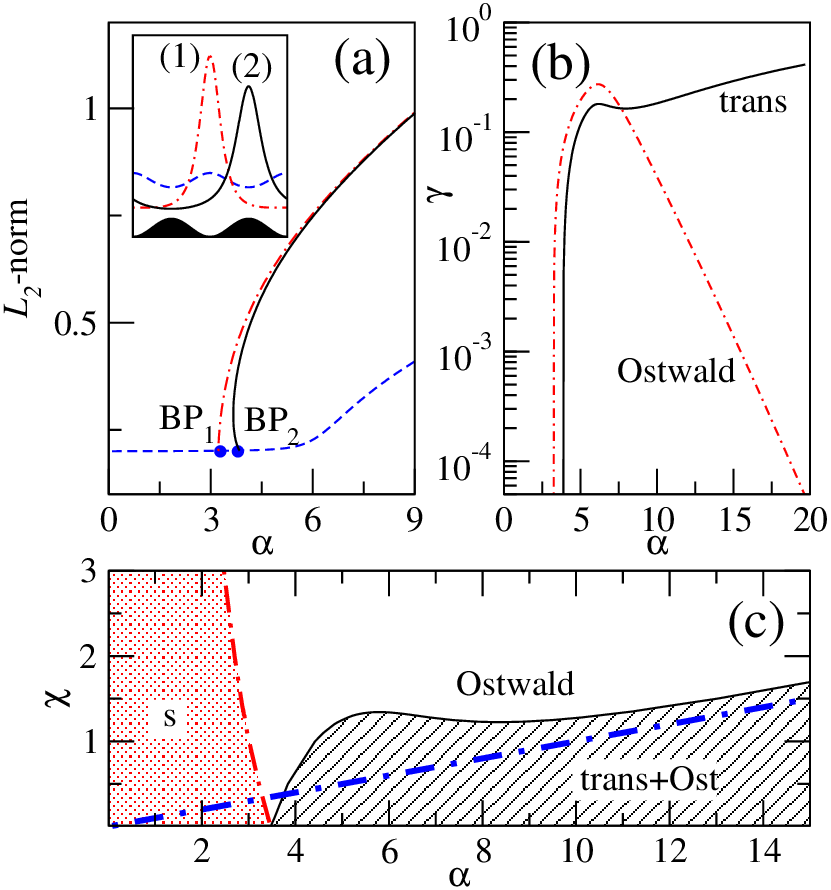}
\caption{(Color online)  
\label{F4} (a) The $L_2$-norm of the steady state solutions of
Eq.\,(\ref{FP}) for $\chi=0.5$, $\rho_0=1/L$ and $S=2L=10$. The dashed
line corresponds to period-$L$ ($L=5$) solutions, the  dot-dashed line and the solid line correspond
to period-$2L$ solutions. Two branches of period-$2L$ solutions emerge
at the pitchfork bifurcations ${\rm BP}_1$ and ${\rm BP}_2$. The inset
shows the channel potential (shaded) and three density profiles: the
period-$L$ solution (dashed line) obtained for $\alpha=3.5$, the
period-$2L$ solution on the first branch at $\alpha=3.4$ (dot-dashed
line), and the period-$2L$ solution on the second branch
(solid line) at $\alpha=3.8$. (b)
The two leading eigenvalues $\gamma$ of the period-$L$ solutions: the
Ostwald mode and the translation mode (trans). All other parameters
are as in (a). (c) {Stability diagram} of the two coarsening
modes in the $(\chi,\alpha)$ plane. Period-$1L$ solutions are stable
in the shaded area marked by (s). The translation mode is stable above
the solid line. The Ostwald mode (Ost) is always unstable {outside the
shaded area.} The thick dot-dashed line corresponds to the estimate for the stability threshold of the translation mode given by Eq.\,(\ref{vol_stab}).}
\end{figure}

First, we consider the most simple case of particles with $\sigma=0$ --
i.e.\ with no hard-core interaction. We set $\rho_0=1/L$ and $S=2L$,
corresponding to just one particle per period-$L$, {i.e., we
  expect `clusters' with either one or two particles.}
Fig.\,\ref{F4}(a), shows the $L_2$ norm of the steady states of
Eq.\,(\ref{FP}) as a function of $\alpha$, obtained for $L=5$. We see
that the period-$L$ solution, { which for $\alpha\lesssim6$
  corresponds to a weakly modulated density distribution (see inset
  Fig.\,\ref{F4}(a)), undergoes two primary pitchfork bifurcations:
  one at $\alpha \approx 3.3$, and the other one at
  $\alpha\approx3.8$, where two new branches of steady state
  period-$2L$ solutions emerge, as shown by the red dot-dashed line
  and by the black solid line in Fig.\,\ref{F4}(a). The first one,
  bifurcating at ${\rm BP}_1$ consists of stable period-$2L$
  solutions, that represent strongly localised two-particle clusters
  centered at the minimum of $U(x)$.  The branch that emerges at ${\rm
    BP}_2$ consists of unstable localised two-particle clusters
  (unstable period-$2L$ solutions) centered at the maximum of $U(x)$
  (see inset of Fig.\,\ref{F4}(a)).}

At each pitchfork bifurcation the period-$L$ becomes unstable with
respect to one of the coarsening modes, i.e., for $\alpha\gtrsim3.8$,
the period-$L$ solution is unstable with respect to the two coarsening
modes described above. At the first branching point ${\rm BP}_1$ the
system becomes unstable to the Ostwald ripening mode. Indeed, the
centers of the two period-$L$ clusters induced by the potential $U(x)$
are located in the minima of $U(x)$. Therefore, the position of the
clusters does not change as one switches to the first branch. In
contrast, when moving from the period-$L$ solution onto the second
branch (above its saddle-node bifurcation) the position of clusters
changes from the minimum to the maximum of $U(x)$, clearly indicating
that the second branching point ${\rm BP}_2$ corresponds to the onset
of the translation mode of coarsening. Note that at large $\alpha$ the
two branches of solutions practically coincide.

The two leading eigenvalues $\gamma$ of the steady period-$L$
solutions are shown in Fig.\,\ref{F4}(b). They become positive at the
respective pitchfork bifurcation. For $\alpha$ not too far above these
thresholds, the Ostwald mode (dot-dashed line) dominates over the
translation mode (solid line). However, as $\alpha$ is further
increased, the eigenvalue of the Ostwald mode rapidly decreases,
whereas the eigenvalue of the translation mode increases almost
linearly with $\alpha$. This shows that for strongly attractive
particles, the translation mode dominates the cluster coalescence
process.

The stability diagram in Fig.\,\ref{F4}(c) shows the locus of the two
branching points ${\rm BP}_1$ and ${\rm BP}_2$ in the $(\alpha,\chi)$
plane. The red dot-dashed line marks the first
bifurcation of the period-$L$ state, associated with the onset of the
Ostwald mode. The  black solid line corresponds to the
onset of the translation mode. {The left hand shaded region
  corresponds to linearly stable period-$L$ state.  Note that for any
  fixed $\alpha$ one can stabilize} the translation mode by an
external potential $U(x)$ with a sufficiently large amplitude
$\chi$. Remarkably, however, the Ostwald mode cannot be stabilized for
any choice of $\chi$, implying that two period-$L$ clusters will
always eventually merge as a result of the diffusion process that is
the basis of the Ostwald coarsening mechanism.  We emphasize though
that the characteristic time scale $\tau \sim 1/\gamma$ associated
with the Ostwald mode diverges {exponentially for large $\alpha$
  as can be clearly seen from Fig.~\ref{F4}(b), where $\gamma$ is
  plotted on a logarithmic scale.} This means that for strongly
attractive particles the transfer of material from one cluster to the
other via the Ostwald mode is a very slow process, as one would expect
for particles diffusing over a high potential barrier.

\begin{figure}[t]
\centering
\includegraphics[width=0.99\columnwidth]{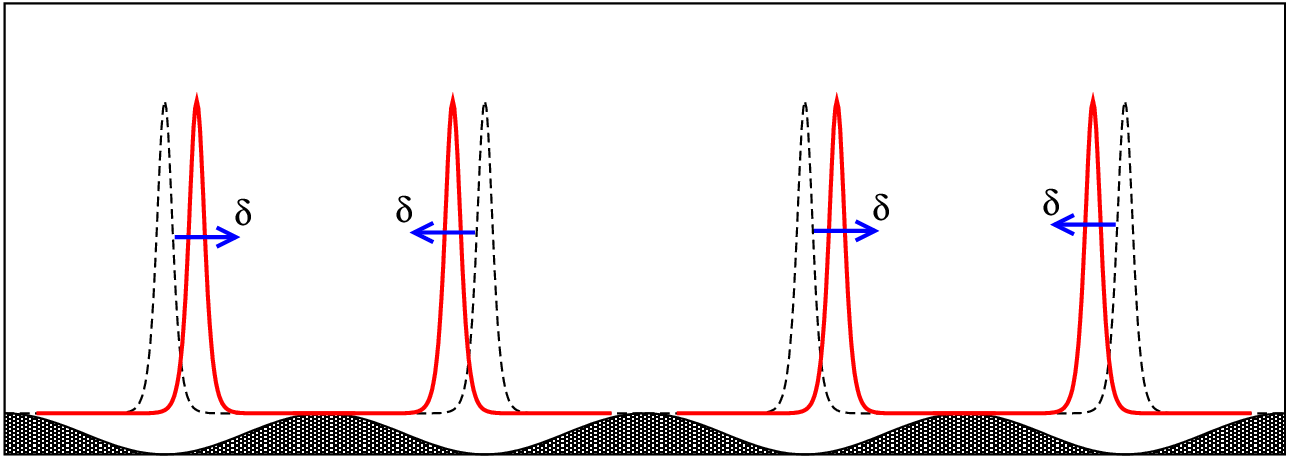}
\caption{(Color online)  
\label{F5} Schematic representation of the translation coarsening mode. The initial clusters (dashed lines) move by a distance $\delta$ towards one of their neighbors. The external potential $U(x)$ is shown by the shaded area. }
\end{figure}

One can give a rather simple qualitative estimate for the stability
threshold of the translation mode (dashed line in
Fig.\,\ref{F4}(c)). This can be done by relating the peaks in the
density profile to individual particles. Thus, we associate the
period-$L$ cluster with $N=\rho_0 L$ particles located at the minima
of $U(x)$, as shown by the dashed lines in Fig.\,\ref{F5}. Since the
attraction strength between particles decreases exponentially with
distance, we only need to take into account the interactions between nearest neighbour particles (clusters). When two neighbouring clusters move towards each other by a distance $\delta$, then the total interaction energy $E(\delta)$ per cluster can be written as $E(\delta)=\rho_0 L\left[-\alpha e^{-(L-2\delta) }+ 0.5(U(x_{\rm min}+\delta)+U(x_{\rm min}-\delta))\right]$. By setting $E''(0)=0$, we find that the equilibrium configuration $\delta=0$ is stable with respect to the translation mode when
\begin{eqnarray}
\label{vol_stab}
\pi\chi \geq 2 \alpha L^2 e^{-L}.
\end{eqnarray}
The critical amplitude $\chi$ from Eq.\,(\ref{vol_stab}), is displayed as the dot-dashed line in Fig.\,\ref{F4}(c), which slightly underestimates the true stability threshold (dashed line) because Eq.\,(\ref{vol_stab}) is obtained via purely energetic arguments and does not take into account entropic contributions.

The physical reason why the Ostwald coarsening process for particles with no hard core ($\sigma=0$) cannot be stabilized by increasing the amplitude of the pinning potential $U(x)$ to some threshold value, can be explained as follows: Particles with no hard core repulsion can always be squeezed into a cluster of arbitrarily small size. So, initially, strong periodic potentials $U(x)$ split the particles into dense clusters, each located at a local minimum of $U(x)$. However, even in the limit $\chi \rightarrow \infty$, the question of the stability of the dense clusters that are located in the minima of $U(x)$, becomes the standard escape problem of a single particle from a potential well. The effective energy barrier that a given particle must overcome in order to escape a cluster is given by the combination of the pinning potential and the energy associated with the attraction to the cluster. It is well-known that the escape rate always remains non-zero, no matter how high is the potential barrier. This ensures that the temperature driven Ostwald mode diffusive process remains active even for very strongly pinning potentials $U(x)$. Intuitively, however, one may expect this mechanism to break down in the case where the particles do have a hard core ($\sigma>0$), as these can no longer be squeezed into a cluster below the size of $\sigma N$.

\subsection{Stabilization of clusters by strongly corrugated potentials}
\begin{figure}[t]
\centering
\includegraphics[width=0.99\columnwidth]{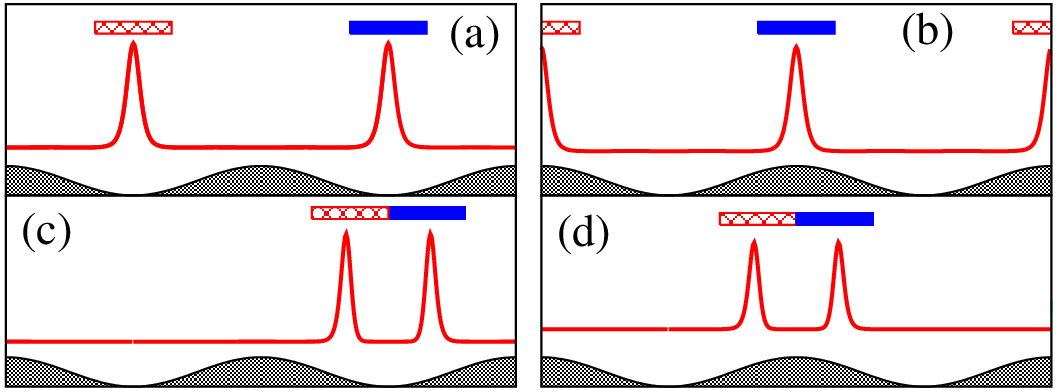}
\caption{(Color online)  
\label{F6} Schematic diagram of the steady state solutions (solid lines) of Eq.\.(\ref{FP}) with $N=2$ strongly attracting hard rods with $\rho_0=1/L$ and $S=2L$. The shaded area represents $U(x)$. The rod length is $\sigma=1.5$.}
\end{figure}

Under similar conditions, hard rods with non-zero length $\sigma\not=0$
behave qualitatively different from the point-like purely attracting
particles considered in Sec.\,\ref{secIIIB}. Even an arbitrarily small
$\sigma$ has a dramatic effect on the stability diagram displayed in
Fig.\,\ref{F4}(c). For any $\sigma\neq0$, the Ostwald mode can be
stabilized by a sufficiently strong corrugation potential $U(x)$.
{Note that, in the large $\chi$ regime the translation mode is
  also stable as shown for $\sigma=0$ in Sec.\,\ref{secIIIB}.  As a result,
  for any $\sigma\not=0$ a strong pinning potential can stabilize clusters
  with a} size sufficiently small to fit within a single potential
well of the channel potential.

In order to see why large amplitudes $\chi$ can lead to the
stabilization of clusters, we consider the strong attraction limit
$\alpha \rightarrow \infty$, where the particle picture can be used
for qualitative arguments. If there are only $N=2$ particles within
the system, one finds four different steady state solutions of
Eq.\,(\ref{FP}), as shown schematically in Fig.~\ref{F6}. {The
  two rods are indicated by rectangles and the corresponding density
  distributions are indicated as well. The steady states of these two
  hard rods consists of the individual particles either being at the
  minima (Fig.~\ref{F6}(a)) or maxima (Fig.~\ref{F6}(b)) of the
  channel potential $U(x)$, or they can form a compact two-particle
  cluster centered at the minimum (Fig.~\ref{F6}(c)) or maximum
  (Fig.~\ref{F6}(d)) of $U(x)$.} Clearly, the steady states (b) and (d)
  are unstable, as any small deviation from the equilibrium will
  result in a translation to an energetically lower state (either (a)
  or (c)).

The energies of the states (a) and (c) are given by $E_a=2U(x_{\rm min})-\alpha e^{-L}$ and $E_c=U(x_{\rm min}+\sigma/2)+U(x_{\rm min}-\sigma/2)-\alpha e^{-\sigma}$, respectively. Consequently, taking into account that $U(x)=-(\chi/2\pi)\cos{2\pi x/L}$, the state (a) corresponds to a global minimum of the energy for $\chi> \pi\alpha (e^{-\sigma}-e^{-L})/(1-\cos{\pi \sigma/L})$. Therefore, when $\chi$ exceeds this threshold value, the clusters with $N=1$ particles, associated with the period-$L$ solutions, are linearly stable.

\begin{figure*}[t]
\centering
\includegraphics[width=1.5\columnwidth]{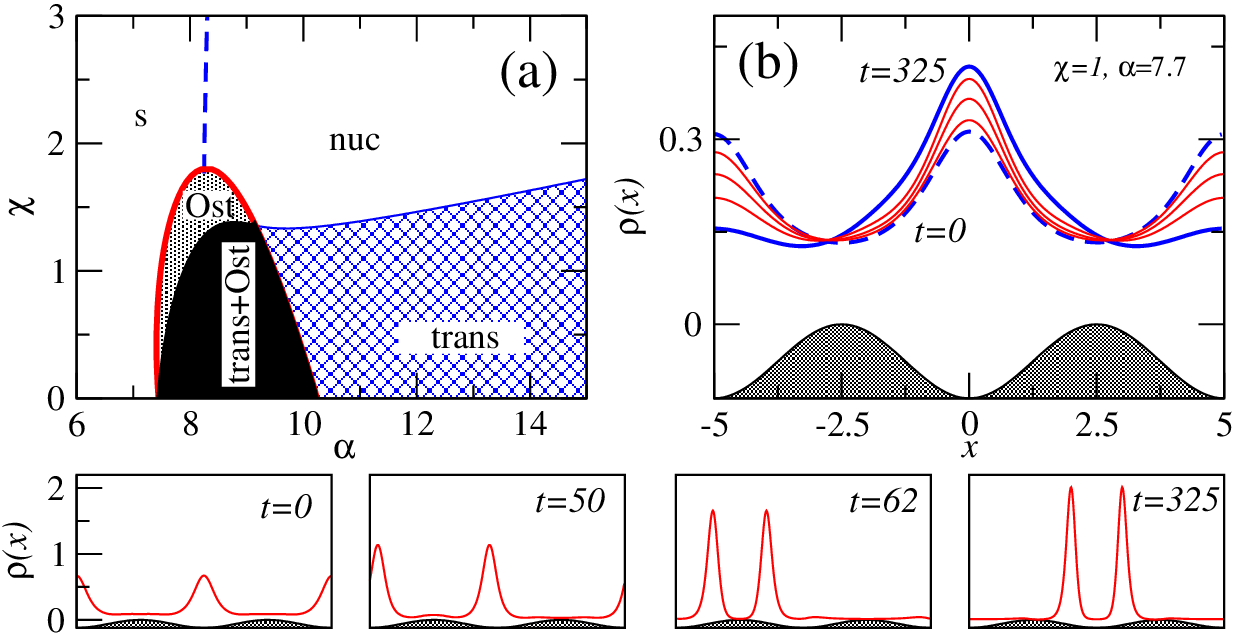}
\caption{(Color online)  
\label{F7} (a) Stability diagram of the two coarsening modes for $\sigma=1.75$ with the remaining parameters as in Fig.\,\ref{F4}(c). (b) The coalescence time evolution of two period-$L$ clusters into a single period-$2L$ cluster mediated by the Ostwald mode, for $\alpha=7.7$ and $\chi=1$. Bottom row: an example of the aggregation dynamics via the translation mode, in this case for $\alpha=10$ and $\chi=1$.}
\end{figure*}

It should be emphasized that the energy-based arguments cannot be used to accurately predict the linear stability threshold. Generally, stabilization occurs at a much lower $\chi$ value, due to the fact that a steady state can be linearly stable even if it has a higher energy than some other steady state. The exact locus of points in the $(\alpha,\chi)$ plane, where linearly stable period-$2L$ solutions exists, is given by the dashed line in Fig.\,\ref{F7}(a), which presents the modification of the stability diagram in Fig.\,\ref{F4}(c) due to a non-zero rod length $\sigma=1.75$. The dashed line marks the locus of the saddle-node bifurcation of the period-$2L$ solutions. 

In the heavy-shaded region of Fig.\,\ref{F7}(a) labelled ``trans+Ost",
both coarsening modes are unstable. The thick red solid line defines
the boundary of the region where the Ostwald mode is unstable. In the
hash-shaded area only the translation mode is unstable, whereas the
Ostwald mode is stable. Outside of the shaded areas the period-$L$
solutions are linearly stable. {However, as the primary
  bifurcations can be subcritical the period-$2L$ clusters can exist
  outside the shaded area. This occurs right of the nearly vertical
  dashed blue line which marks} the saddle-node bifurcation of the
stable period-$2L$ solutions. In this region, labelled ``nuc", the
period-$2L$ clusters must be formed via nucleation. Remarkably, we see
from Fig.\,\ref{F7}(a) that at non-zero $\chi$, one can switch between
the translation mode for coarsening and the Ostwald mode by changing
$\alpha$.

Numerically integrating Eq.\,(\ref{FP}) we obtain the time evolution
of the density profile, with the period-$L$ solution as the initial
condition as displayed in Fig.\,\ref{F7}(b). {We see that for
  $\alpha=7.7$ and $\chi=1$ the dynamics of the coalescence of two
  period-$L$ clusters is indeed via the Ostwald mode.} As expected
from Fig.\,\ref{F7}(a), in this scenario, one of the maxima of the
density profile grows at the expense of the other. Increasing $\alpha$
we find that the translation mode becomes dominant. In the example
displayed along the bottom row in Fig.\,\ref{F7} which is for
$\alpha=10$ and $\chi=1$, we observe a typical coalescence dynamics
due to the translation mode. The two unstable clusters at the
potential minima first move both towards the separating maximum, where
they coalesce into the solution on the BP$_2$ branch. Note that this
solution is itself unstable (on a larger time scale), as it represents
a saddle in function space. However, it acts as an organising center
for the coarsening process as it first attracts the time evolution
along its stable direction (related to its stable eigenvalue closest
to zero) to then expell it along its only unstable direction (in
function space) on a time scale controlled by its unstable eigenvalue.
As a consequence, the fused cluster slowly translates into one of the
potential minima, becoming the solution on the BP$_1$ branch. In other
words, the time scale of the coarsening is controlled by the
eigenvalues of the initial two-cluster state and of the unstable
solution on the BP$_2$ branch. Overall, the eignevalue with the
smallest absolute value gives a good estimate for the time scale of
the coalescence.

\subsection{Larger clusters}

Up to this point we have focused on the coalescence of two `clusters'
which each only consist of $N=1$ particle (on average). The fair question arises as to whether or not the results summarized in Fig.\,\ref{F7} remain at least qualitatively valid for bigger clusters of $N>1$ particles. The complete answer to this question depends on the choice of the length scale parameters such as the rod length $\sigma$, the period $L$ of the channel potential and the system size $S$, all expressed in units of $(\lambda)^{-1}$. Indeed, by increasing the number of particles in the system, i.e.\ by increasing the average density $\rho_0$, one increases the minimal possible size of the cluster, given by $\sigma N$. Clearly, one can expect to uncover a rather different behavior for $\sigma N \ll L$, where the cluster size is much smaller than the period of $U(x)$ and for $\sigma N \sim L$, with cluster sizes comparable to the period of $U(x)$. Furthermore, different dynamical regimes can be expected for different separation distances between the clusters as compared to their size. 

In view of this, it is clear that a complete account of all the
possible dynamical regimes of the cluster formation is beyond the
scope of the present paper. Here we focus on the most simple case
where the smallest cluster size $\sigma N$ is much smaller than both, the
period of $U(x)$ and the separation distance between the clusters. 

\begin{figure*}[t]
\centering
\includegraphics[width=1.5\columnwidth]{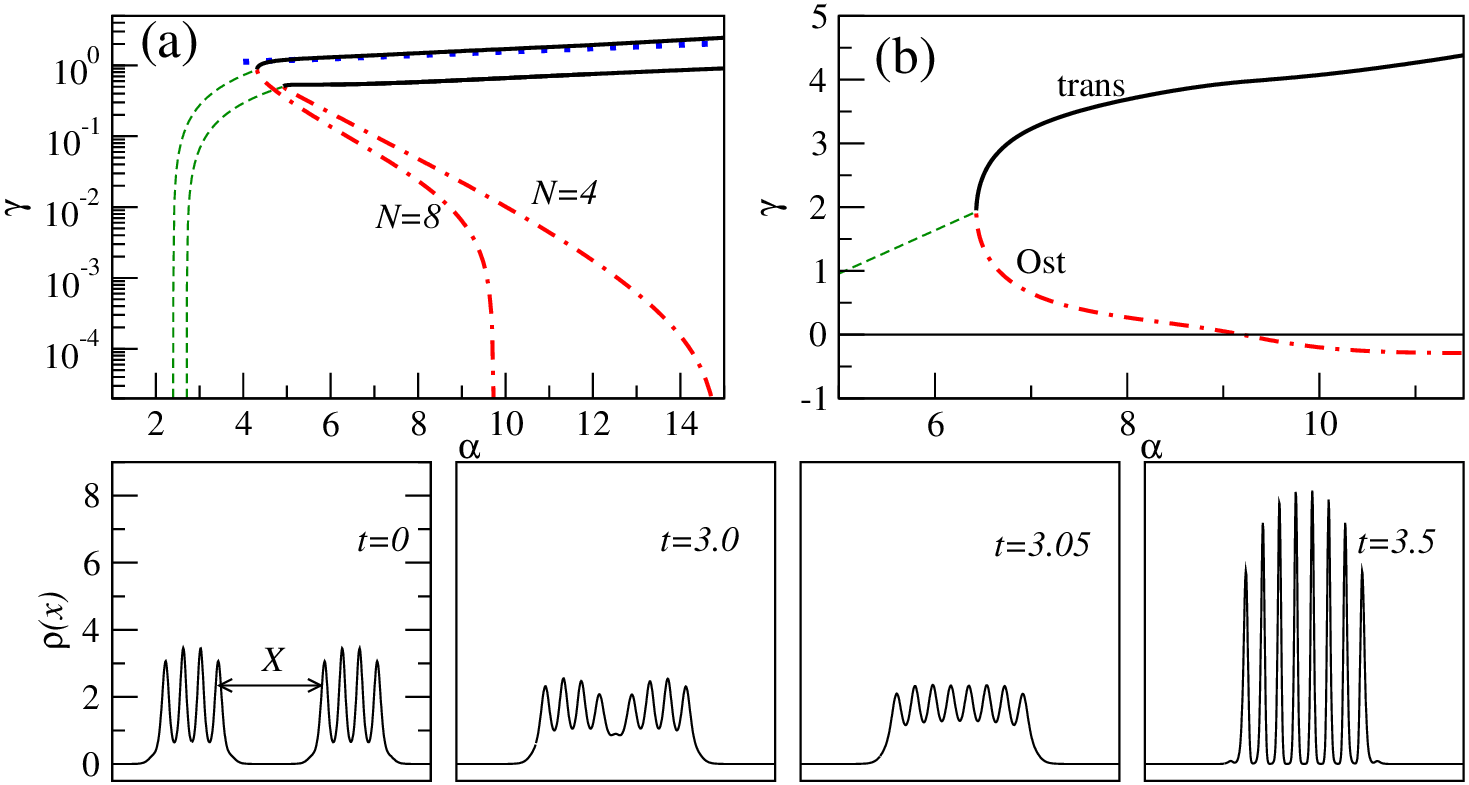}
\caption{(Color online)  
  \label{F8} {Panels (a) and (b) show the dependence on attraction
    strength $\alpha$ of the eigenvalues of the translation (thick
    solid lines) and the Ostwald (dot-dashed lines) coarsening modes
    of period-$L$ solutions in the case of smooth channels ($\chi=0$),
    rodlike particles with $\sigma=0.5$ and domain size $S=2L=10$. The
    thin dashed lines correspond to the growth rate $\beta(k)$ of the
    uniform density state calculated at $k=\pi/5$. (a) gives results
    for densities $\rho_0=0.4$ (i.e.~$N=4$) and $\rho_0=0.8$
    (i.e.~$N=8$), while (b) shows the case $\rho_0=1.2$ (i.e.~$N=12$).
    The thick dotted line in (a) is a linear fit to the increasing
    part of the translation mode eigenvalue. Bottom row illustrates
    the coalescence process for the $\rho_0=0.8$ case in (a) at
    $\alpha=10$, i.e., two $N=4$ clusters coalesce via translation
    into a single $N=8$ cluster.}}
\end{figure*}

Under such conditions, the energy-based arguments used above for the
case of $N=2$, as illustrated in Figs.~\ref{F5} and~\ref{F6}, remain
qualitatively valid for $N>2$. This means that the stabilization of
both coarsening modes by a sufficiently strong potential $U(x)$ can be
expected for larger clusters as well. However, {the presented
  energy-based arguments do not explain a stabilization of the Ostwald
  mode observed even in the case when the channel walls are smooth,
  i.e.\ for $\chi=0$ (see Fig.~\ref{F7}(a) and Fig.~\ref{F8}). In order to understand this
  effect, we fix $=0.5$ and study the dependence of the
  coarsening modes on attraction strength $\alpha$ for
  different numbers of particles in the system as controlled by
  $\rho_0$.}

{Representative examples are shown in Fig.\,\ref{F8}(a) and (b)
  where the eigenvalues of the translation mode and the Ostwald mode
  are denoted by the solid and dot-dashed lines, respectively. Fixing the
  period $L$ and domain size $S=2L$ one notes that the critical value
  of the attraction strenth, at which the Ostwald mode becomes stable,
  depends on the edge-to-edge separation $Y$ between the
  clusters. Indeed, for $N=4$ (i.e., for $S=10$ an edge-to-edge
  separation of $Y\approx L-N\sigma=4$) the stabilization occurs at around
  $\alpha\approx 15$ and for $N=8$ (edge-to-edge separation of
  $Y\approx 3$) it occurs at $\alpha\approx 10$. For an even smaller
  edge-to-edge separation of $Y=2$, the critical attraction strength
  required for stabilization is $\alpha\approx 9$, as shown in
  Fig.\,\ref{F8}(b).  In all cases the eigenvalue of the translation
  mode increases at large $\alpha$ monotonically with $\alpha$.}

\begin{figure}[t]
\centering
\includegraphics[width=0.99\columnwidth]{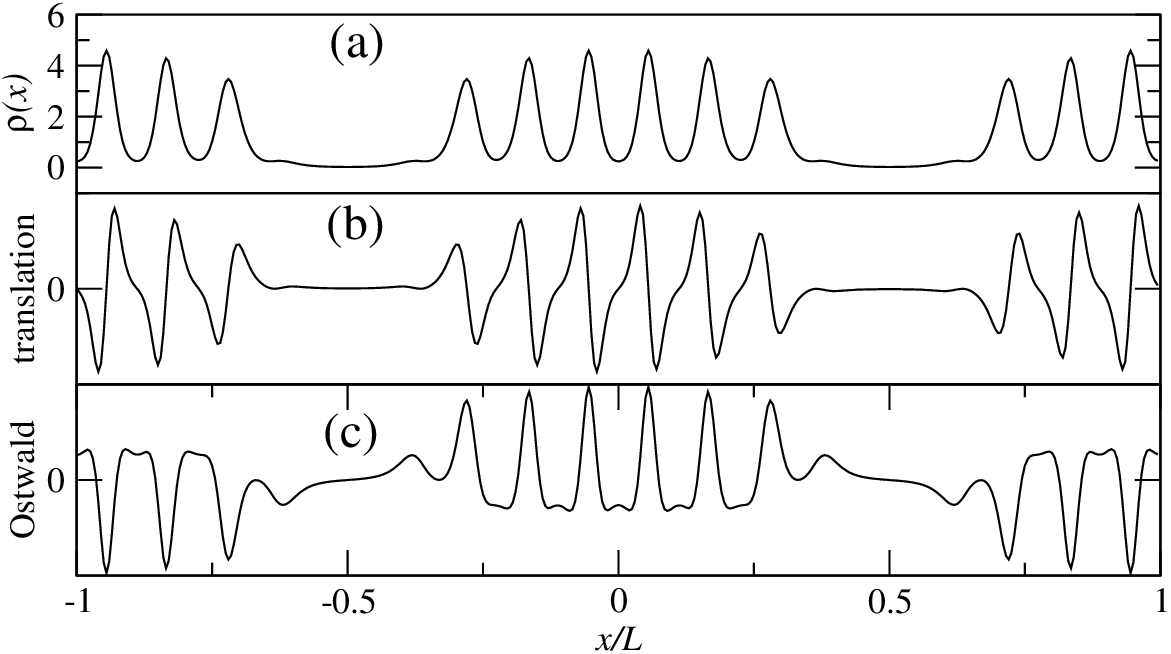}
\caption{(Color online)  
\label{F8a} (a) Steady state period-$L$ density profile for $S=2L$ at $\alpha=10$ and other parameters as in Fig.\,\ref{F8}(b). (b) The unstable translation coarsening mode with eigenvalue $\gamma=4$. (c) The stable Ostwald coarsening mode with $\gamma=-0.2$.}
\end{figure}

The stabilization of the Ostwald mode in the large $\alpha$ limit is
due to the strong attraction that binds the outer particles strongly
to the cluster and so they cannot diffuse away, even under the
influence of a neighbouring cluster. Thus, for coalescence to occur
for large $\alpha$, it must occur via the translation mode. A typical
coalescence scenario of two period-$L$ clusters is {illustrated for
$N=8$ and $\alpha=10$ by the snapshots from a time simulation shown}
in the bottom row of Fig.\,\ref{F8}. At $t=0$ the clusters start to move towards each other due to the dominating unstable translation mode.  As the edge-to-edge distance between the clusters decreases, the strength of the attractive interaction increases exponentially $\sim \alpha \exp{(-Y)}$. Initially, the clusters move towards each other rather slowly, but after a certain time $t=3$, the increasing attractive forces become rather strong and the remaining stages of the coalescence is a very quick process with the typical time scale of $10^{-2}$ as compared to $10^0\dots10^{-1}$ for the initial slow approach. Eventually at $t\approx 3.05$, one larger period-$2L$ cluster is formed with $N=8$ particles. 

For $N=12$, the edge-to-edge separation distance between the
period-$L$ clusters is $Y=L-6\sigma=2 < 6\sigma=3$. The steady state solution as
well as the unstable translation and the stable Ostwald coarsening
modes are shown for $\alpha=10$ in Fig.\,\ref{F8a}. For small
edge-to-edge separations between clusters, i.e. for $Y \ll \sigma N$, the
Ostwald coarsening process is suppressed by the close proximity of the
clusters.
\section{Cross-over between coarsening modes}
\label{Sec:V}
\begin{figure*}[t]
\centering
\includegraphics[width=1.5\columnwidth]{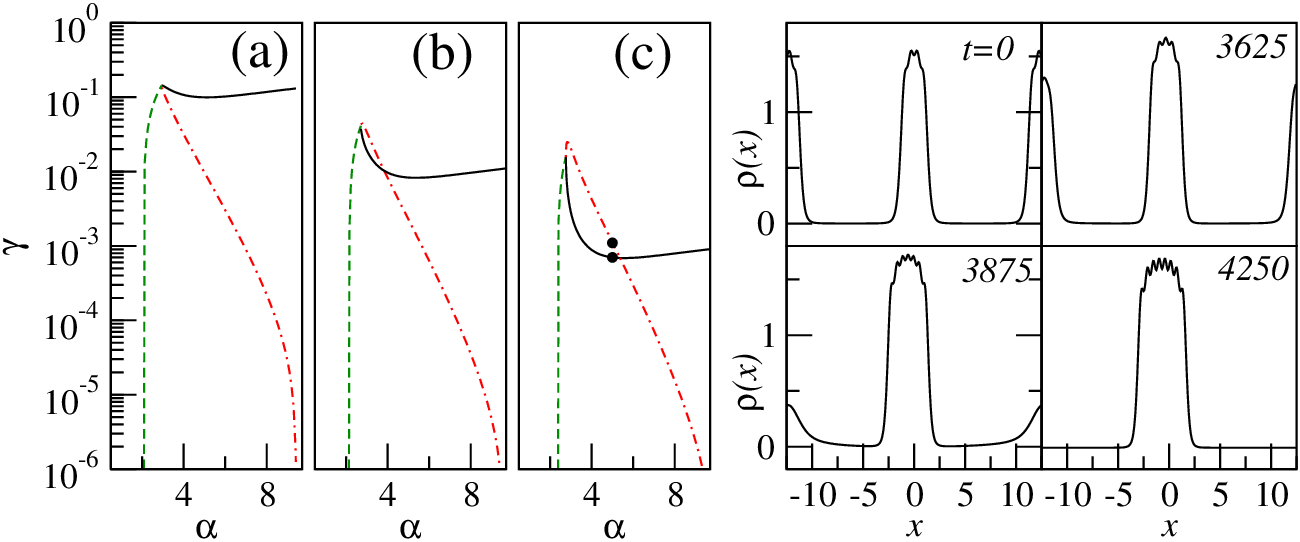}
\caption{\label{F9} (Color online) {Panels (a), (b) and (c) show for
    different cluster distances the two eigenvalues of the coarsening
    modes for a pair of four-particle clusters (period $L=S/2$
    solution) in a channels with smooth walls ($\chi=0$) as a function
    of attraction strength $\alpha$.  The solid lines are for the
    translation mode and the dot-dashed lines for the Ostwald mode.
    The parameters are $\sigma=0.5$ and (a) $S=15$, $\rho_0=8/15$; (b)
    $S=20$, $\rho_0=8/20$; and (c) $S=25$, $\rho_0=8/25$. The filled
    circles in (c) mark the eigenvalues at $\alpha=5$.} The dashed
    lines are obtained from Eq.\,(\ref{disp_rel}) for
    $k=2\pi/S=\pi/L$. Right panel: Time series of density profiles for
    the aggregation of two clusters induced by the dominant Ostwald
    mode, calculated for $S=25$, $\rho_0=8/25$ and $\alpha=5$.}
\end{figure*}

As shown in the previous section, for fixed {center-to-center}
distance between the clusters, the translation mode dominates the
aggregation process for large $N$ and $\alpha$. Remarkably, however,
the time scale associated with the two coarsening modes changes
dramatically, resulting in a {partial} cross-over between the Ostwald and the translation mode, if the separation distance is gradually increased.

{To illustrate what we mean by ``partial cross-over'' between
  the modes, we consider the case with smooth channel walls $\chi=0$
  and fix the number of particles in the system to $N=8$, i.e., we
  look at the coarsening of two clusters of $N=4$ particles each.
  We then follow the branch of period-$L$ solutions in parameter
  $\alpha$ for different separation distances that we control via the
  domain size $S$. This implies that for increasing $S$ we have a
  decreasing $\rho_0=N/S$. The center-to-center distance $X$ between
  the two clusters is then given by $X=S/2$.}

{The two coarsening eigenvalues are shown in
  Fig.\,\ref{F9}(a)-(c) as a function of $\alpha$ for three different
  separations $X$. For $X=7.5$ (Fig.\,\ref{F9}(a)), the translation
  mode dominates in the entire $\alpha$ range where the coarsening
  modes are active ($\alpha \gtrsim 3$) and there is a strong
  separation of the time scales $1/\gamma$ of the two coarsening modes
  beyond the minimum of the eigenvalue of the translation mode at
  $\alpha\approx5$. The Ostwald mode stabilises at
  $\alpha\approx10$. However, at the larger distance $X=10$
  (Fig.\,\ref{F9}(b)), the Ostwald mode dominates the aggregation
  process for a relatively weak attraction $3<\alpha<4$ and a
  cross-over between the coarsening modes occurs at $\alpha \approx
  4$, with the translation mode dominating for stronger attraction. As
  the separation distance $X$ is further increased (see, e.g.~$X=12.5$
  in Fig.\,\ref{F9}(c)), the cross-over point is shifted towards ever
  larger values of $\alpha$.}

An example of the dynamics of coalescence of two $N=4$ clusters is
shown for $S=25$ ($X=12.5$) and $\alpha=5$ in the right panel of
Fig.\,\ref{F9}. There, {the Ostwald mode ($\gamma=0.0011$) acts
  on a shorter time scale $\sim1/\gamma$ than the translation mode
  ($\gamma=0.0007$), as indicated by filled circles in
  Fig.~\ref{F9}(c).  As coarsening progresses, the positions of the
clusters remain practically unchanged. Due to the diffusion of
particles from one cluster to the other that characterizes the Ostwald
mode, one of the cluster shrinks whereas the other one increases in
size. Eventually, one large cluster with $N=8$ particles (period-$2L$
solution) is formed.}

In order to decide which mode is dominant under which conditions, one can use the following qualitative argument, which is valid for strongly attracting particles. 
The time scale associated with the Ostwald mode can be estimated as
follows: The binding energy $w_b$ of a single rod that joins or leaves
the cluster, is independent of the number of particles in the cluster,
provided that this number is rather large.  Therefore, the escape rate
$\nu_\mathrm{esc}$ of the particles at the surface of the cluster is given by
$\nu_\mathrm{esc}\sim \exp{(-w_b)}$. Assuming that only one particle can escape from the cluster at a time, we estimate the characteristic time $t_{\rm Ost}^{\rm esc}$, which is needed for $N$ particles to escape from the cluster
\begin{eqnarray}
\label{ec_mode}
t_{\rm Ost}^{\rm esc}\sim  N \exp{\left(w_b\right)}. 
\end{eqnarray}

After the particle has escaped, it still needs to diffuse the distance $X$ between the two clusters in order to join the other cluster. The diffusion time is estimated as $t_{\rm Ost}^{\rm diff}\sim X^2$.

In order to estimate the time scale associated with the translation
mode, we note that two clusters of $N$ particles each can move and
collide as the result of the attraction between the clusters. This
scenario resembles standard Diffusion-Limited Aggregation
\cite{meakin83,kolb83,weitz84,lin89} with an additional migration of clusters due to attraction.
Thus, if $X \gg \sigma N$ the attraction force between the clusters is proportional to $N^2$, i.e. $f\sim-N^2\,w'(X)$, where $w'(X)$ is the derivative of the pair potential. Therefore, the force per particle is $f/N\sim-N\,w'(X)$. 

The diffusion coefficient $D_N$ of a compact cluster of $N$ particles
is smaller than the diffusion coefficient of a single particle by a
factor of $\sqrt{N}$, i.e. $D_N=1/\sqrt{N}$. {Consequently, the
  time scale of the translation mode is influenced by the time scale
  due to attraction $t_{\rm trans}^{\rm at}$ and the time scale due to
  cluster diffusion $t_{\rm trans}^{\rm diff}$. If they are very
  different the smaller one will dominate.}  In the overdamped limit
we obtain
\begin{eqnarray}
\label{vol_mode}
t_{\rm trans}^{\rm at}=\frac{X\exp{(X)}}{N \alpha },\,\,\,\,t_{\rm trans}^{\rm diff}=\sqrt{N}X^2,
\end{eqnarray}
where we have used the exponential pair potential $w(X)=-\alpha \exp{(-\vert X\vert)}$.

The time scale estimates in Eqs.\,(\ref{ec_mode}) and (\ref{vol_mode})
can now be {compared to the numerical findings reported in
  Figs.~\ref{F8} and \ref{F9}. It is clear that the dependence on the
  cluster size is estimated to $t_{\rm Ost}^{\rm esc}\sim N \gg t_{\rm
    trans}^{\rm diff}\sim \sqrt{N}$, explaining why the aggregation is
  dominated by the translation mode in Fig.\,\ref{F8}. On the other
  hand, when the cluster size $N$ is fixed and $X$ is increased, the
  shortest time scale for the translation mode is ultimately
  determined by clusters diffusion $t_{\rm trans}^{\rm diff} \sim
  X^2\ll t_{\rm trans}^{\rm at}\sim X\exp{(X)}$.} The $X$ dependence
of the Ostwald mode time scale $t_{\rm Ost}^{\rm (diff)}(X)\sim X^2$
is also quadratic. However, for large clusters $N \gg 1$, we have
$t_{\rm Ost}^{\rm (diff)} = t_{\rm trans}^{\rm (diff)}/\sqrt{N}$,
which explains the dominance of the Ostwald mode at large separation
distances as shown in Fig.\,\ref{F9}.


Finally, by examining the dependence on the attraction strength
$\alpha$, we conclude that the Ostwald mode remains effectively frozen
for strong attraction, because the time scale associated with the
escape from the cluster increases exponentially with $\alpha$,
i.e.$t_{\rm Ost}^{\rm esc}(\alpha) \sim e^{\alpha}$.  The translation
mode, on the other hand, increases its dominance $t_{\rm trans}^{\rm
  at} \sim \alpha^{-1}$ due to increased attraction between the
clusters. This linear increase of the translation mode eigenvalue with
$\alpha$ is demonstrated by thick dot-dashed lines in
Fig.\,\ref{F8}(a).
 
{We note that the qualitative estimates derived above cannot be
  used to describe the evaporation-condensation process in the case
  when the edge-to-edge separation distance between two clusters is
  comparable to or smaller than the cluster size. As a consequence,
  the stabilization of the Ostwald mode is not captured by the
  approximation.}
\section{Concluding remarks}
Phase separation occurs in order to minimize the total free energy of the system. A quenched initially uniform system forms clusters which then tend to merge to form bigger aggregates, because this reduces the amount of interfaces between phases and therefore the interfacial contribution to the free energy. The dynamics of the aggregation process, also known as coarsening dynamics, crucially depends on the distance between the clusters, their size and the attraction strength. We have studied the coalescence of two equally-sized clusters separated by a distance larger than the cluster size. The clusters can merge either as the result of their motion and subsequent collision, or as a result of the Ostwald ripening process, whereby particles diffuse from one cluster to another. This process is somewhat akin to an evaporation-condensation process -- i.e.\ the particles evaporate from the smaller drop and then travel to condense on the larger drop. 

We have shown that both coarsening modes can be suppressed and, consequently, the clustering can be `frozen', by a sufficiently strong pinning action of the external potential $U(x)$, associated with the corrugation of the channel walls. In the case of the smooth channel walls, we have demonstrated that the translation mode dominates the aggregation process for large cluster sizes and for strongly attracting particles. In fact, the Ostwald mode can be effectively arrested in the strong attraction limit. Finally, we have demonstrated that there is a cross-over between the coarsening modes, induced by changing the separation distance between the clusters. Regardless of the how strong is the attraction between the particles, the time scale associated with the translation mode can be rendered much larger than the Ostwald mode time scale by increasing the distance between clusters.

Although our results are for a simple 1D model, we believe that our overall conclusions about the behaviour and the interplay of the two different coarsening modes should apply much more generally. We expect two dimensional clusters aggregating on surfaces to behave in a qualitatively similar manner. In this situation the corrugation potential $U(x)$ in our system models a rough potential due to the surface, with the parameter $\chi$ characterising the surface roughness. Similarly, we expect our results to also be relevant to coarsening of clusters in three dimensions. To relate our results to these systems, one should think of the density profiles we calculate as corresponding to the cross-sectional density profile through a pair of two- or three-dimensional clusters. In fact, as mentioned above, because in 1D there is no true phase transition, but in two- and three-dimensions there is a phase transition, the effect of fluctuations that to some extent have been neglected in the present mean-field treatment, will be less influential in two and three dimensions.


\end{document}